\documentclass[%
 preprint,
 amsmath,amssymb,
 aps,
]{revtex4-2}

\usepackage{graphicx}
\usepackage{dcolumn}
\usepackage{bm}
\usepackage{amssymb}
\usepackage{algorithm2e}
\usepackage{amsmath}
\usepackage{xcolor}
\usepackage[colorinlistoftodos,textsize=scriptsize]{todonotes}

\def\tr{{\rm tr}}
\def\bGWH{\boldsymbol{GW}\hspace{-1.4pt}\boldsymbol{H}}
\def\bYlp{\boldsymbol{Y}_{\!\ell p\,}}
\DeclareMathOperator*{\minimize}{minimise~~}

\definecolor{green}{rgb}{0.0, 0.5, 0.0}
\definecolor{green3}{rgb}{0,0.6,0}
\definecolor{indigo}{rgb}{0.3,0.0,0.5}

\begin{document}

\preprint{APS/123-QED}

\title{From STEM-EDXS data to phase separation and quantification using physics-guided NMF}

\author{Adrien Teurtrie}
 \affiliation{Unité Matériaux et Transformations, UMR-CNRS 8207, Université de Lille, Lille, France}
 \affiliation{Electron Spectrometry and Microscopy Laboratory (LSME), Institute of Physics (IPHYS), École Polytechnique Fédérale de Lausanne (EPFL), Lausanne, Switzerland
}
\author{Nathanaël Perraudin}
\author{Thomas Holvoet}%
\affiliation{Swiss Data Science Center (SDSC), ETHZ, Zürich, Switzerland
}%

\author{Hui Chen}
\author{Duncan T.L. Alexander}
\affiliation{Electron Spectrometry and Microscopy Laboratory (LSME), Institute of Physics (IPHYS), École Polytechnique Fédérale de Lausanne (EPFL), Lausanne, Switzerland
}

\author{Guillaume Obozinski}%
\affiliation{Swiss Data Science Center (SDS), ETHZ, Zürich, Switzerland
}

\author{Cécile Hébert}
\affiliation{Electron Spectrometry and Microscopy Laboratory (LSME), Institute of Physics (IPHYS), École Polytechnique Fédérale de Lausanne (EPFL), Lausanne, Switzerland
}

\date{\today}

\begin{abstract}
We present the development of a new algorithm which combines state-of-the-art energy-dispersive X-ray (EDX) spectroscopy theory and a suitable machine learning formulation for the hyperspectral unmixing of scanning transmission electron microscope EDX spectrum images. The algorithm is based on non-negative matrix factorization (NMF) incorporating a physics-guided factorization model. It optimizes a Poisson likelihood, under additional simplex constraint together with user-chosen sparsity-inducing and smoothing regularizations, and is based on iterative multiplicative updates. The fluorescence of X-rays is fully modeled thanks to state-of-the-art theoretical work. It is shown that the output of the algorithm can be used for a direct chemical quantification. With this approach, it is straightforward to include a priori knowledge on the specimen such as the presence or absence of certain chemical elements in some of its phases. This work is implemented within two open-source Python packages, \texttt{espm} and \texttt{emtables}, which are used here for data simulation, data analysis and quantification. Using simulated data, we demonstrate that incorporating physical modeling in the decomposition helps retrieve meaningful components from spatially and spectrally mixed phases, even when the data are very noisy. For synthetic data with a higher signal, the regularizations yield a tenfold increase in the quality of the reconstructed abundance maps compared to standard NMF. Our approach is further validated on experimental data with a known ground truth, where state-of-the art results are achieved by using prior knowledge about the sample. Our model can be generalized to any other scanning spectroscopy techniques where underlying physical modeling can be linearized.
\end{abstract}

\maketitle



\section{Introduction}
\label{sec:intro}

EDX spectroscopy (EDXS) is widely used in analytical electron microscopy. EDXS is a core-level spectroscopy measuring the X-rays emitted by the atoms when they return to ground state after an excitation induced by the incident electron beam in the scanning transmission electron microscope (STEM). In STEM, EDXS performed at high spatial resolution enables determination of the local chemical composition below the nanometric scale \cite{dalfonso_atomic-resolution_2010}. Nevertheless, analyzing materials with complex microstructures and containing multiple phases poses challenges for extracting meaningful information from STEM-EDXS datasets. Three key issues arise when determining chemical composition in such scenarios: noisy spectra due to the low cross-sections for fluorescence yields of X-rays; limitations on exposure time and/or electron beam current before irreversible damage occurs on the sample; and spatial and spectral mixing. Spectral mixing occurs whenever the X-ray peaks of different elements overlap (e.g., Si-K$\alpha$ and Sr-L$\alpha$ lines). There can be spatial mixing along the beam direction whenever the typical scale of the material microstructure is below the sample thickness (i.e.~thickness of tens or hundreds of nanometer). It is then a challenging task to disentangle the contributions of each phase in the spectra. 

To overcome this challenge, the STEM-EDXS data can be modeled as a product of two matrices: one matrix containing the pure spectra of each phase and a second matrix containing their associated spatial distributions. This follows from the principle that each pixel spectrum of the STEM-EDXS datacube can be approximated as a linear combination of the pure phase spectra of the studied material \cite{teurtrie2023espm}. This linear mixing model applies to other fields such as satellite hyperspectral imaging \cite{bioucas-dias_hyperspectral_2012}. Within the latter field, many algorithms have been developed for the decomposition of hyperspectral datacubes, such as non-negative matrix factorisation (NMF) \cite{lee_learning_1999}, independent component analysis (ICA) \cite{hyvarinen_independent_2000}, vertex component analysis (VCA) \cite{nascimento_vertex_2005} or multivariate curve analysis \cite{lavoie_including_2016}. These methods produce a decomposition of the data which makes them adaptable to many different problems, such that, in either direct or modified forms, they have been successfully applied to STEM-EDXS experimental data \cite{lavoie_including_2016, cacovich_unveiling_2018,henry_studying_2020,jany_retrieving_2017,rossouw_multicomponent_2015,shiga_sparse_2016}. However, these algorithms do not benefit from prior knowledge about the shape of the solutions. This can result in a lack of direct interpretability for their solutions, and they have a tendency to overfit and hence fail to remove the noise present in the data.

To solve these issues and improve the decomposition of STEM-EDX spectrum images, here we present a machine learning formulation that uses a physics basis to guide the solutions. The first main contribution of this work is a novel factorization model that integrates the physical modeling of EDX spectra within an NMF framework. This approach has two main additional benefits: the NMF directly outputs a chemical quantification of each phase, and it is possible to introduce prior knowledge on the factors to guide their estimation. Contrary to the spectra, the spatial distribution of phases is not modeled, but it is expected to be smooth, or piecewise smooth, so we also add regularizations in the NMF formulation to avoid overfitting. 

The second main contribution of this paper is the continued development of two open-source Python packages. Previously, we introduced the \texttt{emtables} package for creating X-ray cross-section tables, together with the \texttt{espm} package for the fast simulation of STEM-EDXS datasets \cite{teurtrie2023espm}. Now, we augment \texttt{espm} 
with a dedicated NMF algorithm for the decomposition of STEM-EDXS data that incorporates designed constraints, regularization, and a physics-informed factorization model that makes recourse to the \texttt{emtables}-generated X-ray cross-sections. \texttt{espm} is thus updated into a versatile toolbox for the analysis of EDX spectrum image. These Python packages are freely available and provide the user with advanced control over the functionality of the algorithm and simulations.

In the following, we first describe the physics-informed factorization model. Then, we present an efficient algorithm to estimate pure phase spectra from STEM-EDXS datasets. The effectiveness of our method is tested on both synthetic and experimental datasets. This is described in the subsequent parts of this paper, where we detail the parameters of dataset simulation, the experimental methods, and the results obtained on synthetic and experimental datasets.

\section{The factorization model}
\label{sec:fact_model}

In STEM-EDX spectrum imaging, the electron beam is scanned over an area of the sample, which leads to the emission of X-rays. The raw data comprises X-ray counts measured by the detector at each location of the beam and at various energy levels. The scan is usually performed in a grid pattern and each position of the beam corresponds to a pixel in the data. The count distribution follows a Poisson distribution with the mean (and rate) equal to the magnitude of the spectrum at that specific energy level and location.
We start from the premise that the spectrum of a column of material corresponds to the weighted sum of pure phase spectra.
This leads to the formulation of the following factorization model for the data matrix $\boldsymbol{Y} = (y_{\ell, p}) \in \mathbb{R}^{L \times P}$:
\begin{equation}
\label{eqn:basic_decomp}
    \boldsymbol{Y} \approx \boldsymbol{D} \boldsymbol{H},
\end{equation}
where ``$\approx$'' comes from the Poisson noise. Here, $\boldsymbol{D} = (\delta_{\ell, k}) \in \mathbb{R}^{L \times K}$ is the matrix whose columns are the spectra of pure phases and $\boldsymbol{H} = (h_{k, p}) \in \mathbb{R}^{K \times P}$ is the matrix whose columns are the spatial abundances (i.e.~proportion of each phase) at each pixel. $L$ is the number of energy channels, $P$ is the number of pixels and $K$ is the expected number of phases in the observed sample. This linear mixing model has been widely utilized as a basis to process EDXS data~\cite{bioucas-dias_hyperspectral_2012}.

Without additional constraint, there is an infinite number of pairs $(\boldsymbol{D}, \boldsymbol{H})$ that are solutions to equation \eqref{eqn:basic_decomp}, and in most cases a significant fraction of the elements of $\boldsymbol{D}$ and $\boldsymbol{H}$ take negative values. Hence, in general, the obtained $\boldsymbol{D}$ cannot be directly interpreted as a collection of spectra with physical meaning.

To obtain physically meaningful solutions, some formulations further constrain $\boldsymbol{D}$ and $\boldsymbol{H}$. For instance, NMF constrains $\boldsymbol{D}$ and $\boldsymbol{H}$ to be non-negative and ICA constrains the rows of $\boldsymbol{H}$ to be statistically independent. While these formulations are applicable to a broad range of problems, they still lack sufficient constraints to guarantee that the resulting decomposition matches the permissible spectral features dictated by the physics of EDXS.

We therefore propose a formulation which is much more structured and constrained. First, we propose to further decompose $\boldsymbol{D}$ as the product $\boldsymbol{G} \boldsymbol{W},$ so that the factorization model becomes:  
\begin{equation}
\label{eqn:GW_decomp}
    \boldsymbol{Y} \approx \bGWH 
\end{equation}
where $\boldsymbol{G} = (g_{\ell, m}) \in \mathbb{R}^{L \times M}$ is a pre-computed design matrix whose columns correspond to the modeled X-ray emissions. $M$ is the maximal number of pure spectral features in modeling each EDX spectrum. $\boldsymbol{W} = (w_{m, k}) \in \mathbb{R}^{M \times K}$ is a matrix of learnable parameters whose $k$\textsuperscript{th} column represents the amount of $\boldsymbol{G}$'s spectral feature in the $k$\textsuperscript{th} phase. We note that in~\eqref{eqn:GW_decomp}, the product $\bGWH$ is only an approximation to $\boldsymbol{Y}$. First, the observed X-ray emissions conform to a Poisson distribution parametrized by $\bGWH$. This is detailed in Section~\ref{sec:optimization}. Second, the model relies on the thin foil approximation for the calculation of the absorption \cite{teurtrie2023espm}, which can lead to errors of up to a few percent \cite{hsu2022accurate} in typical samples $\sim$100 nm thick. The matrix $\boldsymbol{H}$ plays the same role here as in~\eqref{eqn:basic_decomp}.

We now provide details on the design of the matrix $\boldsymbol{G}$ so that the product $\boldsymbol{GW}$ models the shape of the EDX spectra of pure phases. The modeling of X-ray emission mainly applies to STEM-EDXS and is based on our earlier work~\cite{teurtrie2023espm}. The EDX spectrum is the sum of two contributions: 

\begin{itemize}
    \item The characteristic X-ray emissions result from discrete transitions originating from inner-shell ionization of the probed atoms by the electron beam. The energies of such transitions are specific to each chemical element. For a given element $\mathsf{Z}$, the X-ray emissions in the $\ell$\textsuperscript{th} energy channel can be modeled as: 
    \begin{equation}
    \label{eqn:carac_xrays}
        \kappa_{\ell,\mathsf{Z}} = \sum_{\mathsf{i},\mathsf{j}} x_{\mathsf{ij}}(\mathsf{Z}) \, \frac{1}{\Delta(\varepsilon^{\mathsf{ij}}(\mathsf{Z})) \sqrt{2\pi} } \, \exp{^{-\frac{1}{2} {\left( \frac{\varepsilon_\ell - \varepsilon^{\mathsf{ij}}(\mathsf{Z})}{\Delta(\varepsilon^{\mathsf{ij}}(\mathsf{Z}))} \right)}^2}}  
    \end{equation}
    In this formulation, $\varepsilon$ is the energy of the measured X-ray, $\Delta$ is the energy broadening of the detector, and $x_\mathsf{ij}$ is the X-ray emission cross-section of the transition between the $\mathsf{i}$\textsuperscript{th} sub-shell and the $\mathsf{j}$\textsuperscript{th} sub-shell. The values of $\Delta$ should be calibrated by the experimentalist. The effects of absorption and detection efficiency are included in the calculation of $x_\mathsf{ij}$, as described in our previous publication \cite{teurtrie2023espm} along with more details about our modeling.

    \item The bremsstrahlung X-ray emissions result from a momentum loss of the electrons within the electric field of the probed material and produce a continuous spectrum. This radiation is commonly parametrized as a polynomial of the energy \cite{chapman_understanding_1984}. However, such polynomials can potentially take negative values, which is not possible physically. To address this issue, we propose a parameterization of an appropriate form of non-negative quadratic polynomial, via non-negative coefficients which guarantee that the polynomial remains positive, as follows:  

    \begin{flalign}
    \label{eqn:brstlg_xrays}
    \begin{split}
    b_{\ell} & = \left(\gamma_{0} \, \frac{\tilde{\varepsilon}_\ell}{\varepsilon_{\ell}} \left( 1 - \tilde{\varepsilon}_\ell \right) + \gamma_1 \, \frac{\tilde{\varepsilon}_\ell^{2}}{ \varepsilon_{\ell}}\right) \, \mathsf{d}_\ell(\varepsilon_\ell) \, \mathsf{c}_\ell(\varepsilon_\ell) \\
    & = \gamma_0 \, b_{\ell}^{(0)} + \gamma_1 \, b_{\ell}^{(1)}    \\
    & \text{with} \quad \tilde{\varepsilon}_{\ell}=\frac{e_0 - \varepsilon_{\ell}}{e_0} ,
    \end{split}
    \end{flalign}
    where $\gamma_0$ and $\gamma_1$ are the parameters of the model, $e_0$ is the energy of the beam and $\mathsf{d}_\ell(\varepsilon_\ell)$ and $\mathsf{c}_\ell(\varepsilon_\ell)$ are the detection efficiency and absorption correction, as calculated in \cite{teurtrie2023espm}, respectively.  See Appendix~\ref{sec:brems_appendix} for more details. With this parameterization, the bremsstrahlung can be modeled by two numbers ($\gamma_0$ and $\gamma_1$) for each phase.
    
\end{itemize}

The composition of a single phase can be expressed with vector $\boldsymbol{c} = (c_{\mathsf{z}})$ in atomic percentages where $\mathsf{z} \in \{\mathsf{Z}_1, \dots, \mathsf{Z}_{M'}\}$ are the corresponding atomic numbers, with $M'  = M-2 $. The intensity at the $\ell$\textsuperscript{th} energy channel $\phi_{\ell k}$ of the $k$\textsuperscript{th} phase can thus be written as  \cite{watanabe_quantitative_2006}: 
\begin{align}
\begin{split}
\label{eqn:EDX_model}
    \phi_{\ell k} & = N_A \, N_e\, I_e \, \tau \, \Omega \, \rho^{*}_k \, \sum_{\mathsf{z}} c_{\mathsf{z} k}  \, \kappa_{\ell, \mathsf{z}} + b_{\ell k} \\
    & = \xi_k \left( \, \sum_{\mathsf{z}} c_{\mathsf{z} k}\,  \kappa_{\ell, \mathsf{z}} + \frac{b_{\ell k}}{\xi_k} \right),
\end{split}
\end{align}
where $N_A$ is the Avogadro number, $N_e$ is the number of electrons in a unit of electric charge, $I_e$ is the beam current, $\tau$ is the acquisition time, $\Omega$ is the geometric efficiency of the detector, and $\rho^{*}_k$ is the mass-thickness of the compound. We use the pre-factor $\xi_k$ to isolate the elemental concentrations; it contains the mass-thickness of the k\textsuperscript{th} phase and the other experimental parameters. Eq.~(\ref{eqn:EDX_model}) can thus be factorized as
\begin{equation*}
\label{eqn:vec_product_EDX}
\begin{split}
\boldsymbol{\phi}_k & = \xi_k
\begin{pmatrix}
    \kappa_{1,\mathsf{Z}_{1}} & \dots & \kappa_{1,\mathsf{Z}_{M'}} \\
    \vdots & \ddots & \vdots  \\
    \kappa_{L,\mathsf{Z}_{1}} & \dots & \kappa_{L,\mathsf{Z}_{M'}} \\
\end{pmatrix}
\begin{pmatrix}
    c_{\mathsf{Z}_{1},k} \\
    \vdots \\
    c_{\mathsf{Z}_{M'},k} \\
\end{pmatrix} \\
& + \xi_k
\begin{pmatrix}
    b_1^{(0)} & b_1^{(1)} \\
    \vdots & \vdots \\
    b_L^{(0)} & b_L^{(1)} \\
\end{pmatrix}
\begin{pmatrix}
    \gamma_{0,k}/\xi_k \\
    \gamma_{1,k}/\xi_k \\
\end{pmatrix} \\
& = \xi_k \left( \boldsymbol{K}\boldsymbol{c}_k  + \boldsymbol{B}\boldsymbol{\gamma}_k \right).
\end{split}
\end{equation*}
Often, in samples studied using STEM-EDXS, several phases coexist in the acquired spectrum image and their proportions vary across the scanned area. To reflect this, we further develop Eq.~(\ref{eqn:EDX_model}) by introducing the abundances of each phase at each pixel of the scan $\eta_{k, p}$:
\begin{flalign*}
\begin{split}
\label{eqn:matr_product_EDX}
\boldsymbol{\Phi} & = \left[
\begin{pmatrix}
    \kappa_{1,\mathsf{Z}_{1}} & \dots & \kappa_{1,\mathsf{Z}_{M'}} \\
    \vdots & \ddots & \vdots \\
    \kappa_{L,\mathsf{Z}_{1}} & \dots & \kappa_{L,\mathsf{Z}_{M'}} \\
\end{pmatrix}
\begin{pmatrix}
    c_{\mathsf{Z}_{1}, 1} & \dots & c_{\mathsf{Z}_{1}, K} & \\
    \vdots & \ddots & \vdots \\
    c_{\mathsf{Z}_{M'},1} & \dots & c_{\mathsf{Z}_{M'},K} \\
\end{pmatrix}
\right. \\
& + \left. \begin{pmatrix}
    b_1^{(0)} & b_1^{(1)} \\
    \vdots & \vdots \\
    b_L^{(0)} & b_L^{(1)} \\
\end{pmatrix}
\begin{pmatrix}
    \gamma_{0, 1} / \xi_{1} & \dots & \gamma_{0, K} / \xi_{K} \\
    \gamma_{1, 1} / \xi_{1} & \dots & \gamma_{1, K} / \xi_{K} \\
\end{pmatrix} \right]\\
& \times \begin{pmatrix}
    \xi_{1} \eta_{1, 1} & \dots & \xi_{1} \eta_{1, P} \\
    \vdots & \ddots & \vdots \\
    \xi_{K} \eta_{K, 1} & \dots & \xi_{K} \eta_{K,P} \\
\end{pmatrix} \\
& = \left[\boldsymbol{KC} + \boldsymbol{B\Gamma}\right]\boldsymbol{H}.
\end{split}
\end{flalign*}
The equation above can be further simplified as a single matrix product:
\begin{equation}
\label{eqn:GW_product}
    \boldsymbol{\Phi} = \left(\boldsymbol{KC} + \boldsymbol{B\Gamma}\right)\boldsymbol{H} =
    \left(\begin{bmatrix}
        \boldsymbol{K} & \boldsymbol{B}
    \end{bmatrix}
    \begin{bmatrix}
        \boldsymbol{C} \\
        \boldsymbol{\Gamma}
    \end{bmatrix}\right) \boldsymbol{H}
    = \bGWH.
\end{equation} 

This factorization model has several advantages. Thanks to the modeling of the spectral features, it both decomposes the analysed dataset and denoises its spectral components in one go. Additionally, the decomposition directly provides a chemical quantification from the resulting $\boldsymbol{C}$ matrix (i.e.~a part of the learned $\boldsymbol{W}$ matrix). It should be noted that the quantification is only valid in the case where 
\begin{equation}
    \label{eqn:simplex_const}
    \forall k \quad \sum^{M-2}_{m=1} w_{mk} = 1,
\end{equation}
which is enforced during the learning of $\boldsymbol{W}$. Since there is an arbitrary scaling between $\boldsymbol{W}$ and $\boldsymbol{H}$ when solving Eq.~(\ref{eqn:GW_decomp}), the constraint~(\ref{eqn:simplex_const}) makes the first $M-2$ rows of $\boldsymbol{W}$ correspond to elemental fractions. The last 2 rows of $\boldsymbol{W}$, that correspond to the learning of the bremsstrahlung model, are let free. Therefore, the formalism used in this work makes a direct bridge between the mixed raw data and a chemical quantification, phase by phase. One should note that a more physically accurate model of the EDX spectra would include the spatial variations of the mass-thickness ($\xi_{k,p}$), but this would break the linearity of Eq.~(\ref{eqn:GW_product}), which would in turn greatly complicate the decomposition of experimental datasets. 

Another advantage of this model resides in the possibility of imposing physical constraints on the analysis. This can be done by keeping some elements of $\boldsymbol{W}$ constant during the decomposition. For example, the contribution $w_{\mathsf{z},k}$ of some elements ($\mathsf{z}$) of some phases ($k$) can be kept to $w_{\mathsf{z},k} = 0$. This is particularly useful when it is known that specific element(s) are absent in some phase(s) of the sample but present in other phase(s).

In summary, the factorization model presented here is a tool to improve the NMF decomposition of STEM-EDX spectrum images, such that its results are denoised and quantified. Thanks to this model, physical constraints, based on prior knowledge about the studied material, can be added to further limit the solutions of the unmixing problem. The following section details the optimization process to obtain $\boldsymbol{W}$ and $\boldsymbol{H}$ given $\boldsymbol{Y}$. 

\section{The optimization problem}
\label{sec:optimization}

The goal of the NMF algorithm developed in this work is to find $\dot{\boldsymbol{W}}$ and $\dot{\boldsymbol{H}}$ such that $\boldsymbol{Y} \approx \boldsymbol{G}\dot{\boldsymbol{W}}\dot{\boldsymbol{H}}$.

$\dot{\boldsymbol{W}}$ and $\dot{\boldsymbol{H}}$ are determined through the optimization of a loss function. The loss function is here to measure the discrepancies between $\boldsymbol{Y}$ and the product $\boldsymbol{GWH}$. The goal is then to find $\dot{\boldsymbol{W}}$ and $\dot{\boldsymbol{H}}$ that optimize the value of the loss function and therefore minimise the discrepancies between the data and the model. 

In the most basic formulation of NMF, the matrices $\boldsymbol{W}$ and $\boldsymbol{H}$ are determined by minimizing the square loss function: 

\begin{equation}
    \label{eqn:least_square_loss}
    L_{2}(\boldsymbol{W},\boldsymbol{H}) =  \| \boldsymbol{Y} - \boldsymbol{GWH} \|^2,
\end{equation}

while preserving the following positivity inequalities:

\begin{equation}
    \label{eqn:non-neg_const}
    \forall k,m,p,\quad w_{mk} \geq 0,\: h_{kp} \geq 0
\end{equation}

The optimization of $L_2$ is well-suited when the statistical distribution of the data follows a Gaussian law. In STEM, the inelastic scattering events that lead to X-ray emission, and hence the X-ray emissions themselves, instead follow a Poisson distribution \cite{egerton2011electron}. Given the X-ray detector properties, these statistics are further preserved at the detection level. Thus, the STEM-EDXS data follow a distribution with, according to our model, a mean of: 

\begin{equation}
    \label{eqn:mean_poisson}
    \boldsymbol{\Phi} = (\phi_{\ell,p}),
\end{equation}

where $\boldsymbol{\Phi}'$ corresponds to the modeled spectrum-image associated to the true distribution of materials within the observed sample. Thus, the probability of the observed data ($ \pi_{\phi_{\ell p}}(y_{\ell p})$) follow the expression:

\begin{equation}
    \label{eqn:proba_poisson}
    \pi_{\phi_{\ell p}}(y_{\ell p}) = \frac{(\phi_{\ell p})^{y_{\ell p}}}{y_{\ell p}!}e^{\phi_{\ell p}}.
\end{equation}

Given that this probability of observing the data is parameterized by $\boldsymbol{GWH},$ by the maximum likelihood principle a good way to estimate those parameters is to maximize this Poisson likelihood ($L'_{P})$ with respect to the parameters $\boldsymbol{W}$ and $\boldsymbol{H}$: 

\begin{equation}
    \label{eqn:Poisson_likelihood}
    L'_{P} = \prod_{\ell,p}e^{-(\bGWH_{\ell p})}\: \frac{(\bGWH)_{\ell p}^{\bYlp}}{\bYlp!} 
\end{equation}

This maximization is equivalent to the minimization of the negative log likelihood. We consider the slightly modified objective:

\begin{flalign}
\begin{split}
    \label{eqn:neg_log_poisson_likelihood}
    L_{\text{P}}(\boldsymbol{W},\boldsymbol{H})= \sum_{\ell,p}^{L,P} \left[ - \bYlp\log (\boldsymbol{GWH})_{\ell p}+ (\boldsymbol{GWH})_{\ell p} \right]
\end{split}
\end{flalign}

Lee and Seung proposed efficient optimization algorithms for NMF~\cite{lee_learning_1999,lee_algorithms_2001} using alternating multiplicative updates on $\boldsymbol{W}$ and $\boldsymbol{H}$. If the optimization is begun with non-negative $\boldsymbol{Y}$, $\boldsymbol{W}$ and $\boldsymbol{H}$, non-negativity is preserved by the algorithm updates.

In our formulation of the problem, we additionally enforce both the sum-to-one~(\ref{eqn:simplex_const}) and non-negativity~(\ref{eqn:non-neg_const}) constraints. The linear constraint of Eq.~(\ref{eqn:simplex_const}) is known as a simplex constraint. Furthermore, the simplex and non-negativity constraints on $\boldsymbol{W}$ together induce sparsity (i.e.~induce solutions which have a certain fraction of entries that are exactly equal to zero) in the columns of $\boldsymbol{W}$ \cite{bach_optimization_2011} and, as a consequence, tend to reduce the presence of mixing between the different phases.

The physical model presented above does not put any additional constraint on $\boldsymbol{H}$, except for non-negativity, even though its rows are high dimensional. The fact that the phase composition can be estimated independently at each pixel (i.e.~each sample column) but only from very low counts data would necessarily result in an overfitted and noisy $\dot{\boldsymbol{H}}$ matrix. However, we can exploit two properties of the samples studied in electron microscopy. First, their phase composition will mostly vary smoothly within space. Then, because the samples are very thin, only a very few different phases can be measured at each pixel, with the vast majority of pixels corresponding to only one phase. Thus, the columns of $\boldsymbol{H}$ are typically sparse. In consequence, we propose adding two regularization terms to $L_{P}$. The first term is known as the Laplacian regularization:

\begin{equation}
    \label{eqn:Laplacian_reg}
     r_{\Delta} = \frac{\lambda}{2} \tr\left( \boldsymbol{H} \boldsymbol{\Delta} \boldsymbol{H}^T \right)= \lambda \sum_{k=1}^K \sum_{(n,m) \in E} (h_{kn}-h_{km})^2,
\end{equation}
where $E$ is the set of pairs of (vertically or horizontally) adjacent pixels. In the expression above, $\boldsymbol{\Delta}$ is the $P\times P$ graph Laplacian matrix~\cite{chung_spectral_1997}. The $\tr\left( \boldsymbol{H} \boldsymbol{\Delta} \boldsymbol{H}^T \right)$ is the sum of square differences between neighboring pixels of $\boldsymbol{H}$. 

The more the columns of $\boldsymbol{H}$ are different from their neighbours, the more the value of $r_{\Delta}$ grows. The competition between the value of $r_{\Delta}$ and $L_{P}$ will make $\boldsymbol{H}$ smoother and thus reduce some of the noise.

The second term, which we call logarithmic regularization, is expressed as: 

\begin{equation}
    \label{eqn:log_reg}
    r_{log} = \mu \sum^{K,P}_{k,p} \log \left( h_{kp} + \alpha \right)
\end{equation}

This regularization aims at inducing sparsity in $\boldsymbol{H}$. The sparsity inducing property of the logarithm can be understood as follows. The logarithm is a strictly increasing function, with a steeper slope for small positive values and a lower slope for higher positive values. Therefore, the cost of increasing small $\boldsymbol{H}$ values is higher than the cost of increasing higher values of $\boldsymbol{H}$. This tends to increase the proportion of the predominant phase at each pixel, and thus reduce the mixing between phases. While $\mu$ controls the strength of this regularisation, $\alpha$ controls its slope at $h_{kp} = 0$. 

The combination of smoothness and sparsity using both the Laplacian and logarithm regularizations tends to produce smooth maps and should yield more accurate decompositions. Taking into account the above regularizations and the two constraints (\ref{eqn:non-neg_const}) and (\ref{eqn:simplex_const}), the optimization problem we consider is now: 
\begin{flalign}
\begin{split}
    \label{eqn:full_loss}
    \minimize_{\boldsymbol{W}, \boldsymbol{H}} &  L_{f}(\boldsymbol{W},\boldsymbol{H}) \\ 
    \text{such that}~~ & \boldsymbol{W} \geq \epsilon, \, \boldsymbol{H}\geq \epsilon, \,\sum_{m \leq M-2} \boldsymbol{W}_m = \boldsymbol{1} \\
    \text{and}~~  & L_{f}(\boldsymbol{W},\boldsymbol{H})  = L_{P}(\boldsymbol{W},\boldsymbol{H}) + r_{\Delta}(\boldsymbol{H}) + r_{log}(\boldsymbol{H}),
\end{split}
\end{flalign}
where $\epsilon > 0$ is a small constant close to the machine precision. This constant ensures that the loss function is well-defined across the entire domain. In practical terms, when the value of $\epsilon$ is sufficiently small, the outcomes closely resemble those obtained under a positivity constraint.

Solving the problem formulated above leads to a decomposition of $\boldsymbol{Y}$ that includes the EDX spectroscopy modeling, yielding a non-negative $\boldsymbol{W}$ matrix and a non-negative, row-wise smooth and column-wise sparse $\boldsymbol{H}$ matrix. It is possible to find solutions $\dot{\boldsymbol{W}}$ and $\dot{\boldsymbol{H}}$ for problem \eqref{eqn:full_loss} via an iterative algorithm, as detailed in the next section.

\section{Solving the optimization problem}
\label{sec:updates}

Our method to solve the optimization problem \eqref{eqn:full_loss} is explained in detail in the technical report by Perraudin et al.~\cite{perraudin2024efficient}. The algorithm consists of iterative alternating updates for $\boldsymbol{W}$ and $\boldsymbol{H}$. These updates are a generalisation of the multiplicative update from Lee and Sung~\cite{lee_algorithms_2001}. Generally, the multiplicative updates present the advantage of being fast~\cite{hien_algorithms_2021} while naturally preserving the non-negativity constraint. Indeed, it is sufficient to initialize the algorithm with positive $\boldsymbol{W}$ and $\boldsymbol{H}$ for constraint (\ref{eqn:non-neg_const}) to be respected throughout the optimization process. The two multiplicative updates derived in this work come from Proposition 1 and eq. (22) in~\cite{perraudin2024efficient}. We refer the reader to Appendix~\ref{app:updates} for details on how these updates are derived for this contribution. The resulting updates are formulated as follows:

\begin{flalign}
    \label{eqn:update_W}
    \begin{split}
        \boldsymbol{W}^{t+1} & = \mathcal{P}_\epsilon \left( \boldsymbol{W}^t \odot \left(\boldsymbol{G}^{T}\left(\boldsymbol{Y}\oslash \boldsymbol{GW}^{t}\boldsymbol{H}^{t}\right){\boldsymbol{H}^{t}}^{T}\right) \right. \\
        & \left. \oslash \left( \boldsymbol{G}^{T}\mathbf{1}^{L,P}{\boldsymbol{H}^{t}}^{T} + \boldsymbol{u} \boldsymbol{\nu}^T\right) \right) \\
        & =  \mathcal{P}_\epsilon \left( \boldsymbol{W}^t \odot \boldsymbol{\Sigma}^t \oslash \left( \boldsymbol{T}^t + \boldsymbol{u} \boldsymbol{\nu}^T \right) \right)
    \end{split}
\end{flalign}
\begin{flalign}
    \label{eqn:update_H}
    \begin{split}
    \boldsymbol{H}^{t+1} & =  \mathcal{P}_\epsilon \biggl( \boldsymbol{H}^t \\ 
    & \left. \odot \left( {\boldsymbol{W}^{t}}^T \boldsymbol{G}^T \left( \boldsymbol{Y} \oslash \left( \boldsymbol{GW}^t \boldsymbol{H}^t \right) \right) + 8 \lambda \max(\boldsymbol{H})   \right) \right. \\
    & \oslash \left. \left( {\boldsymbol{W}^{t}}^T (\boldsymbol{G}^T \mathbf{1}^{L,P}) + \mu  \oslash (\boldsymbol{H}^t + \alpha) \right. \right. \\
    & \left. + 8 \lambda \max (\boldsymbol{H}) + \lambda \boldsymbol{H} \boldsymbol{\Delta}   \right) \Bigr) \biggr),
    \end{split}
\end{flalign}
where the $\oslash$ and $\odot$ symbols represent element-wise matrix division and multiplication, respectively. The function $\mathcal{P}_\epsilon(x) = \max(x, \epsilon)$ corresponds to the projection on the convex set $x\geq \epsilon$. For each element $x$ of $\boldsymbol{H}$ of $\boldsymbol{W}$, it assigns $\epsilon$ if $x<\epsilon$ and $x$ otherwise. $\boldsymbol{u}= (u_{m}) \in \mathbb{N}^{M}$ is a vector with $u_m = 1, m \in {1, \dots, M'}$, $u_{M-1} = 0$ and $u_{M} = 0$. To simplify the notation, we define $\boldsymbol{\Sigma} = \left(\boldsymbol{G}^{T}\left(\boldsymbol{Y}\oslash \boldsymbol{GW}^{t}\boldsymbol{H}^{t}\right){\boldsymbol{H}^{t}}^{T}\right) = (\sigma_{mk}) \in \mathbb{R}^{M \times K}$ and $\boldsymbol{T} = \boldsymbol{G}^{T}\mathbf{1}^{L,P}{\boldsymbol{H}^{t}}^{T} = (\tau_{mk}) \in \mathbb{R}^{M \times K}$ as the numerator and a component of the denominator in eq.~(\ref{eqn:update_W}), respectively. The vector $\boldsymbol{\nu} = (\nu_k) \in \mathbb{R}^{K}$ is the Lagrange multiplier used to enforce the sum-to-one constraint (eq.~\ref{eqn:simplex_const}).

After the initialisation of the $\boldsymbol{W}^0$ and $\boldsymbol{H}^0$ matrices, they are updated alternately using the equations (\ref{eqn:update_W}) and (\ref{eqn:update_H}). For the update of $\boldsymbol{W}$, an additional step is required to determine $\boldsymbol{\nu}$ and therefore enforce the sum-to-one constraint. The values of $\boldsymbol{\nu}$ are calculated by solving the following non-linear equation with respect to $\nu_k$:  

\begin{equation}
    \label{eqn:nu_eqn}
    \sum^{M'}_m \max \left( w_{mk}^t \frac{\sigma_{mk}}{\tau_{mk} + \nu_k}, \epsilon \right)  = 1 .
\end{equation}

Given that the left-hand-side of the equation above is an increasing function of $\nu_k,$ the equation can be solved to high-accuracy by dichotomy.

The successive updates are repeated until a convergence criterion is met: either the maximum number of iterations is reached, or the relative variation of the loss function is small enough. A summary of the operation of the algorithm is shown below in Algorithm~\ref{alg:espm}. A detailed proof that the loss function \eqref{eqn:full_loss} decreases under the multiplicative updates \eqref{eqn:update_W} and \eqref{eqn:update_H}) can be found in \cite{perraudin2024efficient}.

\RestyleAlgo{ruled}

\SetKw{Init}{init}{}{}
\SetKwProg{Dichotomy}{dichotomy}{}{}
\SetKwProg{UpdateW}{update W}{}{}
\SetKwProg{UpdateH}{update H}{}{}
\SetKwProg{UpdateG}{update G}{}{}

\begin{algorithm}[hbt!]
\caption{EsPM-NMF${\left( \mu, \lambda, \alpha \right)}$}\label{alg:espm}
\SetKwInOut{Input}{input}\SetKwInOut{Output}{output}
\Input{$K,\boldsymbol{G}, \mu, \lambda, \alpha, tol$}
\Init{$L^0_{f}, H^0\geq 0, W^0\geq 0$}

\While{$\left( L^{t+1}_{f} - L^{t}_{f} \right) > tol$}{
  \UpdateW{}{$\forall m,k$, calculate $ \sigma_{m,k}$ and $\tau_{m,k}$\\
  \Dichotomy{}{$\boldsymbol{\nu} \gets$ equation \ref{eqn:nu_eqn}} 
  $\boldsymbol{W}^{t+1} \gets $ equation \ref{eqn:update_W} }
  
  \UpdateH{}{
  $\boldsymbol{H}^{t+1} \gets$ equation \ref{eqn:update_H}
  }
}
\Output{$\boldsymbol{W}, \boldsymbol{H}$}
\end{algorithm}

In the following, we denote our physics-guided factorization algorithm, that is implemented in the $\texttt{espm}$ package, as ESpM-NMF. The parameters $\mu$, $\epsilon$ and $\lambda$ are considered as hyperparameters of the decomposition; in our notation, they are specified in the order ESpM-NMF$\left( \mu, \lambda, \alpha \right)$.


\section{Materials and methods}
\label{sec:matandmethods}

Two Python libraries were developed. The first, \texttt{emtables}\footnote{publicly available at https://github.com/adriente/emtables}, calculates the characteristic X-ray emission cross-sections. It produces the tables for the data simulations. The second library, \texttt{espm}\footnote{publicly available at https://espm.readthedocs.io/en/latest/}, creates the simulated data, the matrix $\boldsymbol{G}$ and runs the decomposition algorithm. The structure of the algorithm is based on the scikit-learn framework \cite{buitinck_api_2013,pedregosa_scikit-learn_2011}, while the user interface relies on the hyperspy library \cite{de_la_pena_hyperspyhyperspy_2021}.

To test the effectiveness of the ESpM-NMF$\left( \mu, \lambda, \alpha \right)$ algorithm -- both against other factorization algorithms and in function of chosen hyperparameters and constraints -- simulated data were produced using our previously-developed data simulation method \cite{teurtrie2023espm}. Given that this provides a ground truth to compare against, this therefore allows algorithmic performance to be quantified. The advantages of using our framework to simulate data are twofold: the shape of the ground truth exactly matches with the columns of $\boldsymbol{G}$, and a full control of the simulation is kept.

The simulated data are inspired by a contemporary geology problem: the mineralogy of Earth's deep mantle, where the phases Bridgmanite (Brg), Ferropericlase (Fp), and Ca-perovskite (Ca-Pv) are believed to predominate. The spectra associated to each phase of this problem are the same as the ones that were simulated in our previous work \cite{teurtrie2023espm}. The abundances of the data in \cite{teurtrie2023espm} were based on chemical maps of experimental STEM-EDX spectrum images. In the present study, we have taken a different approach, employing small spherical nanoparticles to represent the abundances of Fp and Ca-Pv, while the abundance of the Brg is set to be the matrix that surrounds the particles. The goal of this distribution of phases is to simplify the interpretation of the results of the decomposition. The maximum proportion of the nanoparticles is 0.5 so that there is always a mixing between Brg and the other minerals. This way, it remains challenging to unmix the phases.

All the produced datasets are of dimension $P_x \times P_y \times L = 128 \times 128 \times 1980$. The signal-to-noise ratio (SNR) for a signal that follows a Poisson distribution is $\sqrt{N}$ where $N$ is the intensity of that signal. In this work, we therefore use the total number of counts per spectrum as a simplified indicator of the SNR of the simulated datasets. Two types of datasets are produced: low SNR datasets and high SNR datasets with 18 and 293 counts per spectrum, respectively. The values of $N$ are chosen so that they correspond to a data acquisition of 10 minutes with 500 counts.s$^{-1}$ or 8000 counts.s$^{-1}$ on the detector. For each SNR value, four datacubes were created, each having an identical ground truth but with a different Poisson sampling.

We test the performance of ESpM-NMF against three other algorithms, that themselves are implemented outside of \texttt{espm}: NMF, ICA and VCA-SUNSAL. The scikit-learn implementations of both NMF and ICA are used. For NMF, the optimization is done using multiplicative updates to minimise the negative log likelihood loss function \cite{lee_algorithms_2001}. For ICA, a principal component cnalysis (PCA) pre-processing step is necessary. PCA is performed using the normalization of Keenan et al.~\cite{keenan_accounting_2004} to account for the Poisson noise. Then, the data reconstructed with the first three components are fed to the ICA algorithm. The ICA is applied on the data using the fastICA algorithm \cite{hyvarinen_independent_2000}. For VCA-SUNSAL, the reconstructed spectra are obtained using VCA \cite{nascimento_vertex_2005}, while the maps are determined using SUNSAL \cite{bioucas-dias_alternating_2010} based on the results of VCA. The different algorithms are each applied to all the low and high SNR simulated datasets. For the decomposition of high SNR data, the ESpM-NMF$\left( \mu, \lambda, \alpha \right)$ and NMF algorithms are initialized using non-negative double singular value decomposition \cite{boutsidis_svd_2008} with zero values filled with small random numbers. For low SNR data, the ESpM-NMF$\left( \mu, \lambda, \alpha \right)$ and NMF algorithms are initialized at random. These initializations were chosen to obtain the best optimization results. For ICA, ESpM-NMF$\left( \mu, \lambda, \alpha \right)$ and NMF, the same random seed is chosen for the initialization. The version of VCA that is tested in this paper has its own non-random initialization method.

For the results of these algorithms to be compared to the ground truth, their output is first renormalized. The normalized $\boldsymbol{D^*}$ and $\boldsymbol{H^*}$ are obtained as follows: 

\begin{align*}
    \label{eqn:norm_results}
    \boldsymbol{D^*} & = \boldsymbol{D} \boldsymbol{S}^{-1}\\
    \boldsymbol{H^*} & = \boldsymbol{S} \boldsymbol{H},
\end{align*}

where $\boldsymbol{S} \in \mathbb{R}^{K \times K}$ is a diagonal matrix that scales $\boldsymbol{D}$ and $\boldsymbol{H}$ so that $\sum_k \boldsymbol{D^*} \approx 1$.

Although visual inspection of the spectra produced by the decomposition may be useful, a metric is necessary to quantify the results. In this paper, two metrics are used to measure the quality of the reconstructed spectra and abundances. For spectra, the angle between the decomposition results and the ground truth is calculated as: 

\begin{equation}
    \label{eqn:angle}
    \alpha (\boldsymbol{d}_1, \boldsymbol{d}_2) = \arccos{\left(\frac{\boldsymbol{d}_1 \cdot \boldsymbol{d}_2}{\|\boldsymbol{d}_1\| \, \|\boldsymbol{d}_2\|}\right)},
\end{equation}
where $\boldsymbol{d}_1,\boldsymbol{d}_2 \in \mathbb{R}^L$. These vectors correspond to columns of $\boldsymbol{D}$, where typically $\boldsymbol{d}_1$ and $\boldsymbol{d}_2$ would correspond to the spectrum of a true phase and to the spectrum of a reconstructed phase, respectively. An angle of 0 deg represents a perfect agreement between the two spectra.
For abundance maps, the mean squared errors (MSE) between the reconstructed maps and the true maps are calculated. For each decomposition, the results are quantitatively compared with the ground truth using the mean MSE for the phase maps and the mean angles for the phase spectra.

Following the tests on synthetic data, we demonstrate the validity of ESpM-NMF$\left( \mu, \lambda, \alpha \right)$ on experimental data by applying it to data collected by Rossouw et al. \cite{rossouw_multicomponent_2015}. In their work, they did two STEM-EDXS experiments: one on core-shell nanoparticles with an FePt core and a Fe$_3$O$_4$ shell and a second one on the FePt bare cores. They were able to show a successful unmixing of the core from the shell using ICA, and they could verify their unmixing procedure using the results on the bare cores. In our work, we use both the bare cores and core-shell data, that are publicly available in the \texttt{hyperspy-demos} repository \footnote{link to the github repository: https://github.com/hyperspy/hyperspy-demos} \footnote{direct dropbox link to the data:\\ https://www.dropbox.com/s/ecdlgwxjq04m5mx/HyperSpy\_demos\_EDS\_TEM\_files.zip?raw=1} to test the ESpM-NMF$\left( \mu, \lambda, \alpha \right)$ algorithm.


\section{Results and discussion}
\label{sec:resanddiscussion}

We now test the effectiveness and accuracy of the ESpM-NMF$\left( \mu, \lambda, \alpha \right)$ algorithm on simulated data, and also compare the quality of decomposition results to those from NMF, ICA and VCA. The series of tests on high SNR data is used to verify whether it is possible to recover the ground truth. On the series with a low SNR, the limits of the decomposition algorithms presented here are tested.

A total of six analyses are performed on each dataset. The first three decompositions are made using VCA, ICA and NMF, which together represent the current state of the art. The remaining decompositions are all made using ESpM-NMF$\left( \mu, \lambda, \alpha \right)$, applying different parameters to investigate the effects of the EDXS physical modeling, additional constraints, and regularizations on the decomposition. The first test is performed without additional hyperparameters, and so is labeled ``ESpM-NMF$\left( 0, 0, 0 \right)$". In some experimental cases, the absence of certain elements in some phases can be deduced from the physical conditions under which the sample was obtained. This is for instance the case for the Fp phase that we simulate here, which cannot contain Si. In this case, the relevant elements of $\boldsymbol{W}$ can be set to 0.0 for each phase; this configuration is labeled ``ESpM-NMF$\left( 0, 0, 0 \right)$ fW(0conc)". The ESpM-NMF$\left( \mu, \lambda, \alpha \right)$ algorithm is also tested without prior knowledge but using predefined regularizations; these tests are labeled ``ESpM-NMF$\left( 0.004, 1, 0.01 \right)$".

\begin{figure}
    \centering
    \includegraphics[width=\textwidth]{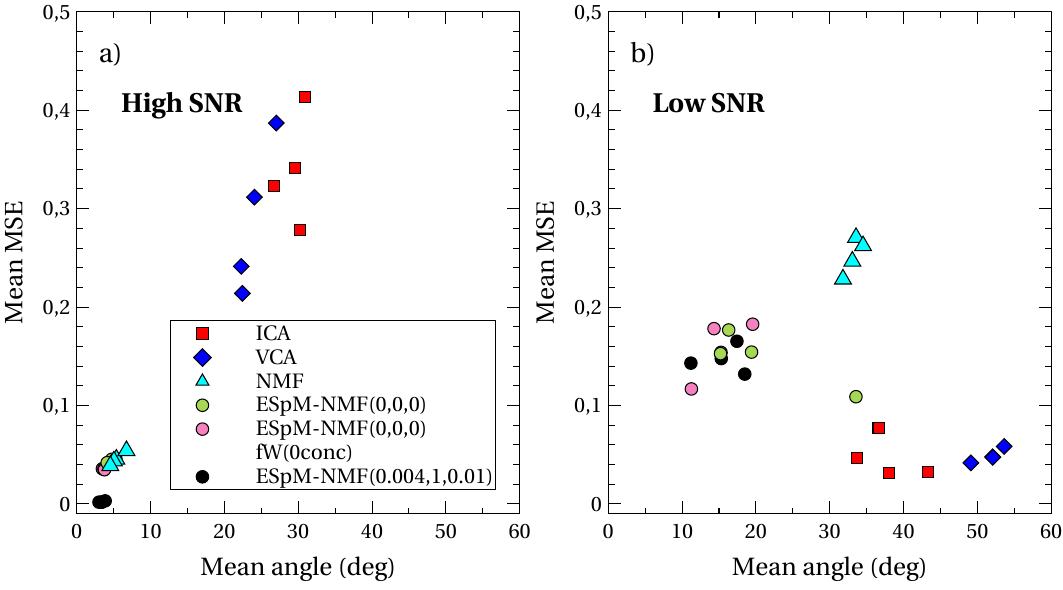}
    \caption{a) mean MSE as a function of the mean angle for all the tested algorithms for high SNR datacubes. b) mean MSE as a function of the mean angle for all the tested algorithms for low SNR datacubes. For each algorithm there are four data points, one for each Poisson sampling. }
    \label{fig:angles_vs_mse}
\end{figure}

Figure \ref{fig:angles_vs_mse} plots the mean MSE of the maps versus the mean angles of the spectra for all the analyses. In these plots, the bottom left corner represents the best results, closely matching the ground truth; it is the opposite case for the top right corner. 

The results on the high SNR data (figure \ref{fig:angles_vs_mse} a)) are discussed first. Both ICA and VCA perform systematically worse than the other algorithms, giving mean angles above 20 deg and mean MSE above 0.2. ``Standard'' NMF produces more accurate results, with mean angles between 4.5 and 6.7 deg and mean MSE between 0.038 and 0.054. The results of ESpM-NMF$\left( 0, 0, 0 \right)$ are very similar, with mean angles between 4.1 and 5.3 deg and mean MSE between 0.040 and 0.046. This behaviour can be understood by the fact that NMF and ESpM-NMF$\left( 0, 0, 0 \right)$ share very similar optimization methods, except for the sum-to-one constraint of the latter on $\boldsymbol{W}$. It also implies that the incorporation of physical modeling into the decomposition has little influence on average spectral angles in this high SNR case. Nevertheless, later we show that it does help eliminate artifact noise peaks in individual phase spectra solutions. Adding constraints by applying the ESpM-NMF$\left( 0, 0, 0 \right)$ fW(0conc) configuration leads to an even better agreement, with the mean angles ranging from 3.5 to 3.8 deg and mean MSE ranging from 0.035 to 0.036. The added prior knowledge also tends to stabilize the solutions, since there is a lower spread between the data points of the different synthetic datasets. The ESpM-NMF$\left( 0.004, 1, 0.01 \right)$ configuration exhibits the best results, with mean angles between 3.0 and 3.8 deg and with a tenfold improvement in mean MSE, that ranges from 0.002 to 0.003.

\begin{figure}
    \centering
    \includegraphics[width=\textwidth]{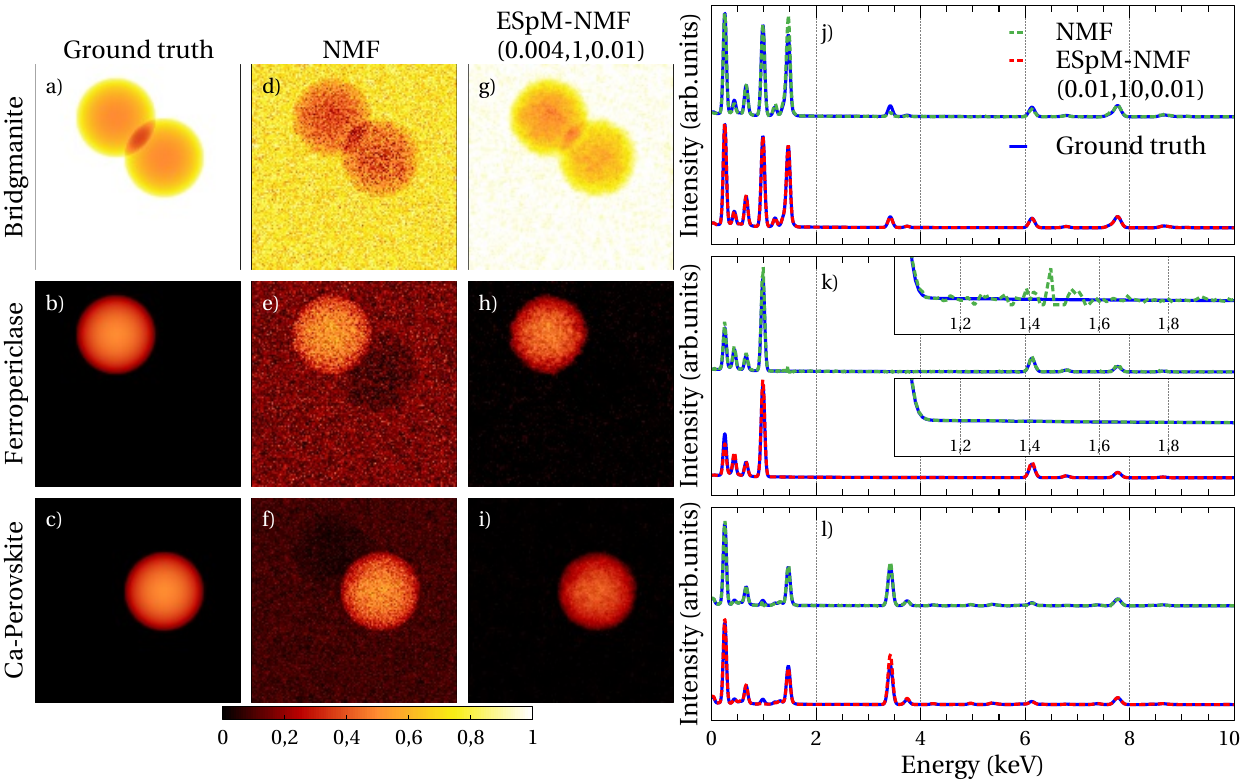}
    \caption{Typical results on a high SNR sample. a), b), c) Ground truth values of the abundance maps of Brg, Fp and Ca-Pv, respectively. d), e), f) Abundance maps resulting from the NMF decomposition. g), h), i) Abundance maps resulting from the ESpM-NMF$\left( 0.004, 1, 0.01 \right)$ decomposition. j), k) and l) Comparison between the ground truth spectra of Brg, Fp and Ca-Pv and the results of the unmixing algorithms, respectively. In k), the top and bottom insets correspond to the same zoom around the peak of Al-K$\alpha$ as reconstruted using NMF and using ESpM-NMF$\left( 0.004, 1, 0.01 \right)$, respectively. }
    \label{fig:res_image_N293}
\end{figure}

To look more closely at this effect of regularization, Figure \ref{fig:res_image_N293} shows a comparison between the ground truth and the NMF and ESpM-NMF$\left( 0.004, 1, 0.01 \right)$ decompositions for high SNR data. It can be clearly seen from Figure \ref{fig:res_image_N293} d) and e) that, even though the three phases are identified by ``standard'' NMF, there is still some mixing, mainly between the Brg and Fp. On the contrary, for ESpM-NMF$\left( 0.004, 1, 0.01 \right)$, the Laplacian regularization produces uniform maps and the sparsity induced by the log regularization greatly reduces the level of mixing between the phases, as shown in Figure \ref{fig:res_image_N293} g) and h). The NMF reconstructed spectrum of Figure \ref{fig:res_image_N293} k) exhibits a noise-related artifact at 1.5 keV, as highlighted in the inset of Figure \ref{fig:res_image_N293} k). This problem is non-existent for the ESpM-NMF$\left( 0.004, 1, 0.01 \right)$ reconstructed spectrum. Thus, the effect of the physical modeling is similar to an additional noise-reduction filter.

\begin{figure}
    \centering
    \includegraphics[width=\textwidth]{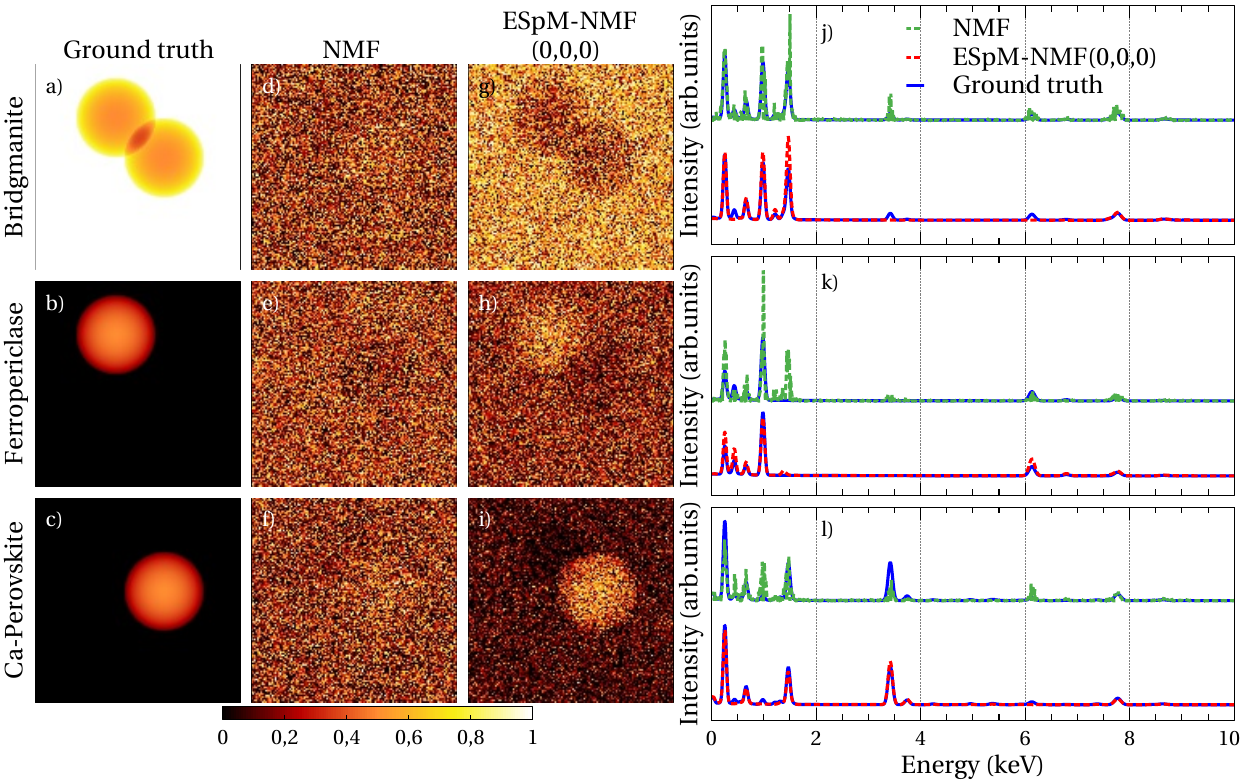}
    \caption{Typical results on a low SNR sample. a), b), c) Ground truth values of the abundance maps of Brg, Fp and Ca-Pv, respectively. d), e), f) Abundance maps resulting from the NMF decomposition. g), h), i) Abundance maps resulting from the ESpM-NMF$\left( 0, 0, 0 \right)$ decomposition. j), k) and l) Comparison between the ground truth spectra of Brg, Fp and Ca-Pv and the results of the unmixing algorithms, respectively.}
    \label{fig:res_image_N18}
\end{figure}

With the weak signal of the low SNR data, it is clear from the angles and MSE in Figure \ref{fig:angles_vs_mse} that no algorithm achieves results as close to the ground truth as for high SNR data. Both ICA and VCA show high mean angles above 30 deg. Paradoxically, their mean MSE are low, with values below 0.1. However, these low mean MSE hide the fact that the ICA and VCA algorithms do not actually identify the presence of the three discrete phases. Rather, they each give one abundance map that contains all three phases, with ones everywhere, and two other abundance maps with zeroes everywhere. In comparison, the NMF analysis gives a cluster of mean angles just above 30 deg, and mean MSE above 0.2. Nonetheless, the use of NMF does not result in a precise identification of the phases, as the abundance maps are predominantly randomly populated.

Compared to ICA, VCA and NMF, Figure \ref{fig:angles_vs_mse} b) shows that ESpM-NMF gives a markedly improved performance on the low SNR data. All three variants exhibit approximately the same performance as each other, with comparatively low mean angles ranging from 11 deg to 20 deg and mean MSE between 0.11 and 0.18. ESpM-NMF$\left( 0, 0, 0 \right)$ shows one outlier at 33.6 deg; we believe that this derives from the random initialisation not being fully reliable. Here, we choose to emphasize the effects of physical modeling on the decomposition results and resulting discrepancies, which we do by comparing the results of NMF and ESpM-NMF$\left( 0, 0, 0 \right)$ in Figure \ref{fig:res_image_N18}. This figure shows that, for the low SNR data, it is  extremely challenging to identify the presence of several phases using NMF. The mixing between the different phases in the three components is so strong that there is no unique way to match them to the ground truth phases. On the contrary, using the ESpM-NMF$\left( 0, 0, 0 \right)$ algorithm, it is directly possible to recognize the three phases, even without the help of adding prior knowledge or regularizations to the algorithm.

Since the only two differences between ESpM-NMF$\left( 0, 0, 0 \right)$ and NMF are the sum-to-one constraint and the physical modeling, this improvement must derive from one or both of these factors. To find out which, the ESpM-NMF$\left( 0, 0, 0 \right)$ decomposition was tested with $\boldsymbol{G}$ equals the identity matrix, thus removing the modeling of X-rays. A solution very similar to that of NMF was then obtained. The results therefore demonstrate that including physics in the machine learning process is an effective strategy for analysing low SNR STEM-EDXS data. 

\newpage

\begin{figure}
    \centering
    \includegraphics[width=\textwidth/2]{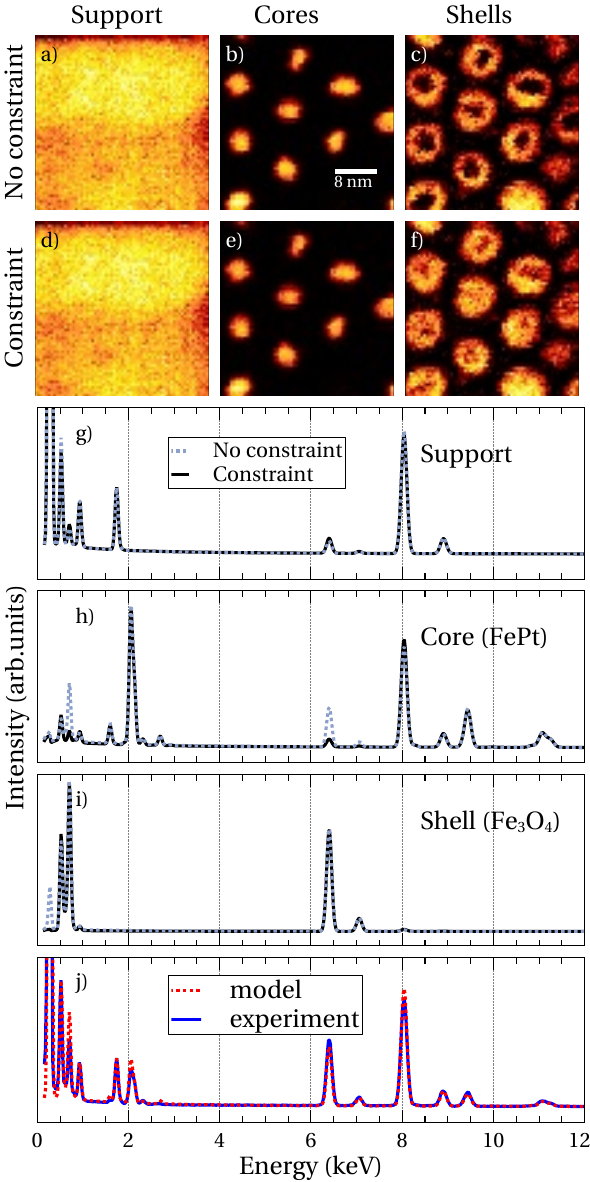}
    \caption{Results of the ESpM-NMF$\left( 0, 0, 0 \right)$ decomposition on experimental data. a), b), c) Abundance maps obtained using the unconstrained ESpM-NMF$\left( 0, 0, 0 \right)$ of the support, cores and shells, respectively. d), e), f) Abundance maps obtained when introducing prior knowledge about the sample, of the support, cores and shells, respectively. g), h), and i) Comparison between the spectra resulting from the constrained (black line) and unconstrained (blue dots) ESpM-NMF$\left( 0, 0, 0 \right)$ decomposition. j) Comparison between the mean of the experimental spectra (blue line) and the mean of the reconstructed spectra (red dots).}
    \label{fig:res_exp}
\end{figure}

\newpage

While special attention is paid to simulating EDXS data as accurately as possible, the simulation cannot fully encompass all of the features of a real experimental dataset. Therefore, we also test the ESpM-NMF$\left( 0, 0, 0 \right)$ algorithm on experimental data. We choose to analyze data on FePt@Fe$_3$O$_4$ nanoparticles collected by Rossouw et al. \cite{rossouw_multicomponent_2015}, since they represent an interesting unmixing problem. Assigning the correct quantities of Fe to the cores and shells is not straightforward in this case. Furthermore, thanks to their analysis of the bare cores, a ground truth is available in Rossouw et al.'s work. This enables us to determine whether our algorithm converges towards physically accurate results, similar to those found in their work that they obtained using ICA.

In their study, Rossouw et al. \cite{rossouw_multicomponent_2015} determined that the bare cores consist, on average, of 18 at. \% of Fe and 82 at. \% of Pt, with a margin of error of $\pm$ 3 at. \%. Additionally, they established that the FePt@Fe$_3$O$_4$ nanoparticles exhibit a core-shell configuration. This configuration implies that, everywhere a particle exists, the proportion of the shell never drops to 0. A successful decomposition using ESpM-NMF$\left( 0, 0, 0 \right)$ should achieve two key results: an accurate composition of the cores and a correct spatial distribution of the shells.

Here, we first perform a decomposition without the use of prior knowledge on the chemical composition of the sample. Its outcomes are shown in Figure \ref{fig:res_exp} under the label ``No constraint''. While the support (Figure \ref{fig:res_exp} a)) is well separated from the particles, the shells (Figure \ref{fig:res_exp} c)) appear perforated. Around the core, some Fe belonging to the shells is instead attributed to the FePt cores, leading to the latter having an overestimated spatial extent and overestimated Fe content. 

This effect can be seen in Figure \ref{fig:res_exp} h), where the Fe K$\alpha$ peak is very intense. Hence, the built-in quantification of the ESpM-NMF$\left( 0, 0, 0 \right)$ decomposition yields a composition of the core of 45 at. \% of Fe and 55 at. \% of Pt. Our method, without external help from the experimentalist, fails to accurately retrieve the true distribution of phases within this dataset. In this configuration, the performance of the ESpM-NMF$\left( 0, 0, 0 \right)$ algorithm is very similar to that of NMF, which also fails to converge towards the correct solution.

Nevertheless, thanks to the modeling presented in section \ref{sec:fact_model}, we can add constraints on the values of $\boldsymbol{W}$ that will guide the decomposition process. To produce our own prior knowledge of the sample, we analyze the bare cores dataset with our method and find that the cores are composed of 16 at. \% of Fe and 84 at. \% of Pt. These proportions are very close to the ratio measured by Rossouw et al.~\cite{rossouw_multicomponent_2015} on both the bare cores and on the core-shell nanoparticles. All the calculations were performed with tabulated theoretical X-ray emission cross sections \cite{teurtrie2023espm} which should ideally be tuned using standards. Small differences in cross section values between our work and those of Rossouw et al.~\cite{rossouw_multicomponent_2015} may well explain the slight discrepancy between our bare core composition and theirs. 

Then, for the next decomposition of the core-shell dataset, we set the values of a part of the elements of $\boldsymbol{W}$ so that the phase corresponding to the cores has a Fe/Pt ratio corresponding to that obtained previously on the bare cores. The results are displayed in Figure \ref{fig:res_exp} under the label ``constraint''. The absence of holes in the calculated spatial distribution of the shells indicates a successful unmixing of the FePt@Fe$_3$O$_4$ nanoparticles dataset. This is confirmed by the reduced intensity of the Fe K$\alpha$ peak in Figure \ref{fig:res_exp} h). To further assess the quality of the decomposition, we compute the chemical composition of the shells. We find a composition of 40 at.~\% of Fe and 60 at. \% of O which is close to the expected 43 at. \% of Fe and 57 at. \% of O of Fe$_3$O$_4$. We obtain this result by adjusting the mass thickness ($\rho^*$) of the model with a thickness of 30 nm and a density of 5.2 g.cm$^{-3}$.

Finally, to ensure that our our constrained modeling approach has not led to any important features of the data being cut during decomposition, we compare the sum of experimental spectra to the sum of the reconstructed spectra. The results are displayed in Figure \ref{fig:res_exp} j), and exhibit a good agreement between the model and the experiment. Therefore, while standard NMF and ESpM-NMF$\left( 0, 0, 0 \right)$ fail to provide a satisfactory unmixing of the FePt@Fe$_3$O$_4$ nanoparticles, by incorporating prior knowledge into the ESpM-NMF$\left( 0, 0, 0 \right)$ decomposition, it is possible to accurately separate the cores from the shells. As well as demonstrating how prior knowledge can greatly enhance the fidelity of decomposition, and together with the algorithm testing on simulated datasets, this result emphasizes the importance of \texttt{espm} providing a versatile decomposition toolbox, whose parameters can be appropriately tuned or optimized for the problem of interest.


\section{Conclusion}

An open source Python software tool called ``\texttt{espm}'' has been developed for the simulation and analysis of STEM-EDXS data. The part of \texttt{espm} responsible for the simulation was described in our previous work \cite{teurtrie2023espm}. In this work, we develop a physics-guided NMF-based algorithm to decompose experimental STEM-EDXS data into phases and their spatial distributions. First, we designed a linear factorization model of STEM-EDXS spectrum images that integrates the physics modeling into NMF analysis. An additional advantage of using this model is that a direct elemental quantification of each phase is inherently made during the decomposition process. We presented here the optimization method to perform the model-based NMF, as derived from the alternating multiplicative updates algorithm of Lee and Seung \cite{lee_learning_1999}. We also added the possibility for the user to apply regularizations to further improve the analysis.

The effectiveness of the developed algorithm was tested on both simulated and experimental data. On the simulated data, it was shown that our method quantitatively outperforms the other tested decomposition algorithms. On data with a high SNR, the ESpM-NMF$\left( 0, 0, 0 \right)$ algorithm shows similar results as the common implementation of NMF but, with the help of regularization, the ESpM-NMF$\left( 0.004, 1, 0.01 \right)$ results match the ground truth almost perfectly. On low SNR data, while standard NMF fails to separate the data into several phases, thanks to the modeling it is possible to identify the different phases using ESpM-NMF. Concerning experimental data, by using constraints on the model, we could successfully reproduce the results of Roussouw et al. \cite{rossouw_multicomponent_2015}. 

NMF is a statistical decomposition method which makes it very flexible, such that it can be applied on a very broad range of subjects. However, since the result is purely statistical, its use requires an interpretation stage. For example, for STEM-EDXS data, standard NMF does not usually produce EDX spectra directly, and at least one quantification step is required. Our approach is to linearize the underlying physical model of EDX and then integrate this linear model into the NMF decomposition. Thanks to this modeling, the decomposition produces directly interpretable spectra, and the model can be constrained to obtain results closer to physical reality. We believe that this linearization method is adaptable to other areas of physics, and hence can be applied to a wide range of decomposition problems.


\section{Acknowledgments}

This project was funded by the Swiss Data Science Center under the SDSC grant C19-07.


\appendix

\section{Parametrization of the breaking radiation.}
\label{sec:brems_appendix}

The modeling of bremsstrahlung has been investigated by several authors; a review of the proposed phenomenological models can be found in the work of Small et al. \cite{small_modeling_1987}. All of the proposed models give a decreasing bremsstrahlung intensity with increasing energy, either with a linear, quadratic or exponential shape. The most recent theoretical work shows that a second order polynomial is sufficient to describe the behaviour of bremsstrahlung for energies above 1 keV \cite{chapman_x-ray_1983, chapman_understanding_1984}. To integrate the bremsstrahlung in the factorization model developed in this work, its parameterization has to be linear and non-negative. 

In the following, we show how the bremsstrahlung equation from \cite{lifshin_use_1974}:
\begin{equation}
\label{eq:bremstrahlung-original}
b'(\varepsilon_{\ell}) = \gamma'_0 \frac{e_0-\varepsilon_{\ell}}{\varepsilon_{\ell}} + \gamma'_1 \frac{ \left(e_0- \varepsilon_{\ell} \right)^{2}}{e_0 \varepsilon_{\ell}}
\end{equation}
can be reparametrized for a new model ($b(\varepsilon_{\ell})$) with only positive coefficients $\gamma_0, \gamma_1$ as used in \eqref{eqn:brstlg_xrays}. 
As the bremsstrahlung is positive (physical constraint), we need to satisfy:
\begin{equation}
\label{eq:bremstrahlung-positivity}
\forall \ell, \, b'(\varepsilon_{\ell}) \geq 0 \hspace{1em} \text{and} \hspace{1em} 0 \leq \varepsilon_{\ell}\leq e_0,
\end{equation}
which puts some constraints on $\gamma'_0, \gamma'_1$. To formulate these constraints, we simplify \eqref{eq:bremstrahlung-original}. 
First, since $\varepsilon_{\ell} \geq 0$, we can multiply $b'(\varepsilon_{\ell})$ by $\varepsilon_{\ell}$ without changing its sign. Second, we introduce the change of variable $\varepsilon'_{\ell} = e_0 - \varepsilon_{\ell}$. We note that $\varepsilon'_{\ell} \geq 0$ for $\varepsilon_{\ell} \in [0, e_0]$. After factorizing $e_0 - \varepsilon_{\ell}$, we obtain:
\begin{equation}
   \dot{b}(\varepsilon'_{\ell}) :=  \frac{\varepsilon_{\ell}}{\varepsilon'_{\ell}} b'(\varepsilon_{\ell}) = \gamma'_0 + \frac{\gamma'_1}{e_0} \varepsilon'_{\ell} \geq 0 \hspace{1em} \forall \ell , \, 0\leq \varepsilon'_{\ell} \leq e_0.
\end{equation}
Since $\dot{b}$ is linear in $\varepsilon'_{\ell}$, ensuring that $ \dot{b}(0) \geq 0,  \dot{b}(e_0) \geq 0$ forces $\dot{b}(\varepsilon'_{\ell})\geq 0$ for any energy $\varepsilon'_{\ell} $ between $0$ and $e_0$. Therefore, we finally obtain two constraints: 
\begin{equation}
\label{eq:constraint-b}
    \gamma'_0 \geq 0 \hspace{1em} \text{and}\hspace{1em} \gamma'_0+\gamma'_1 \geq0.
\end{equation}
from $\dot{b}(0) = \gamma'_0 \geq 0$ and $\dot{b}(e_0) = \gamma'_0 + \gamma'_1 \geq 0$.

Our next step is to search for another parameterisation $\gamma_0, \gamma_1$ such that $\gamma_0^\prime\geq 0, \gamma_1^\prime \geq 0$ for any value of $\gamma'_0,\gamma'_1$ respecting the constraint of \eqref{eq:constraint-b}. Given \eqref{eq:constraint-b}, a trivial solution is
\begin{equation*}
\gamma_1 = \gamma'_0 + \gamma'_1 \geq 0 \hspace{1em}\text{and}
\hspace{1em} \gamma_0 = \gamma'_0 \geq 0.
\end{equation*}
In this new parameterisation, we have $\gamma'_{1}=\gamma_{1} - \gamma_{0}$, which allow us to rewrite \eqref{eq:bremstrahlung-original} as
\begin{equation*}
b'(\varepsilon_{\ell}) 
= \gamma_0 \left( \frac{e_0-\varepsilon_{\ell}}{\varepsilon_{\ell}} - \frac{\left(e_0-\varepsilon_{\ell}\right)^2}{\varepsilon_{\ell} e_0}\right) +\gamma_1 \frac{\left(e_0-\varepsilon_{\ell}\right)^2}{\varepsilon_{\ell} e_0}
\end{equation*}
Without loss of modeling power, we further transform $b'(\varepsilon_{\ell})$ into:
\begin{flalign}
\begin{split}
\label{eq:bremstrahlung-transformed}
b(\varepsilon_{\ell}) & = \frac{b'(\varepsilon_{\ell})}{e_0} \\
& = \gamma_0 \frac{e_0-\varepsilon_{\ell}}{\varepsilon_{\ell} e_0} \left( 1 - \frac{e_0-\varepsilon_{\ell}}{e_0}\right) +\gamma_1 \frac{\left(e_0-\varepsilon_{\ell}\right)^2}{e_0^2 \varepsilon_{\ell}},
\end{split}
\end{flalign}
which gives us the two terms of equation \eqref{eqn:brstlg_xrays}.

For simplicity, we did not address the fact that $ \lim_{\varepsilon_{\ell} \to 0} b(\varepsilon_{\ell})  = \infty$, the previous reasoning could still be established rigorously taking this fact into account.


\section{Updates computation}
\label{app:updates}
In their work, Perraudin et al.~\cite{perraudin2024efficient} introduce novel efficient algorithms for solving the regularized Poisson non-negative factorization problem. The key approach involves the utilization of the block successive upper minimization (BSUM) algorithm~\cite{razaviyayn2013unified}. This algorithm operates by successively minimizing upper bounds of the objective function for each block of variables. Specifically, in the context of regularized Poisson NMF, the two blocks of variables involved are denoted as $\boldsymbol{W}$ and $\boldsymbol{H}$. Ensuring convergence requires that these upper bounds satisfy a crucial property: they must be majorizing functions. Additional details can be found in~\cite{razaviyayn2013unified,perraudin2024efficient}.
To formulate a concrete algorithm, Perraudin et al.~\cite{perraudin2024efficient} construct appropriate majorizing functions tailored to the regularizers employed in their contribution. Subsequently, they compute the minimiser of the majorizing function efficiently, providing a formula for the iterative updates \cite[Proposition 1 and eq. 22]{perraudin2024efficient}. 

The formulation of \cite[eq. 10]{perraudin2024efficient} proposes a minimisation of a general loss function  with the three different type of regularisations, Lipchitz, relatively smooth and concave functions (noted $s_{R}, s_{L}$ and $s_{C}$), as well as linear constraint (noted $\boldsymbol{e}^{T} \boldsymbol{x} = \boldsymbol{1}$). In the present work, the overall loss function we aim to minimise is expressed as follows:
\begin{flalign}
    \label{eq:full_loss_detail}
    \begin{split}
    L_f(\boldsymbol{W}, \boldsymbol{H}) & = {\color{green}L_{\text{P}}(\boldsymbol{W},\boldsymbol{H})} + {\color{orange} r_{\Delta} (\boldsymbol{H})}+ {\color{purple}r_{log}(\boldsymbol{H})} \\
     & = {\color{green} \sum_{\ell,p}^{L,P} \left[ - \bYlp \log (\boldsymbol{GWH})_{\ell p}+ (\boldsymbol{GWH})_{\ell p} \right]} \\ 
    & + {\color{orange}   \frac{\lambda}{2} \tr\left( \boldsymbol{H} \boldsymbol{\Delta} \boldsymbol{H}^T \right)} \\
    & +{\color{purple}  \mu \sum^{K,P}_{k,p} \log \left( h_{kp} + \alpha \right) }
\end{split}
\end{flalign}
Following the nomenclature and colors proposed in \cite{perraudin2024efficient}, we categorize the loss into three distinct parts corresponding to the Poisson loss (in green), a Lipschitz function (in orange), and a strictly concave function (in purple). The Lipschitz and strictly concave functions correspond to $s_{L}$ and $s_{C}$ respectively in \cite[eq. 10]{perraudin2024efficient}. The ESpM-NMF algorithm can be effectively customised with any regularisation that follows the assumption presented in \cite{perraudin2024efficient}.

According to \cite[eq. 16,17,18 of Proposition 1]{perraudin2024efficient}, we derive an update for $\boldsymbol{H}$ as follows:
\begin{flalign}
    \begin{split}
    \boldsymbol{H}^{t+1} & =  \boldsymbol{H}^t \\
    & \odot \left( {\color{green}{\boldsymbol{W}^{t}}^T \boldsymbol{G}^T \left( \boldsymbol{Y} \oslash \left( \boldsymbol{GW}^t \boldsymbol{H}^t \right) \right)} + {\color{orange}8 \lambda \max(\boldsymbol{H})}   \right)  \\
    & \oslash \left( {\color{green}{\boldsymbol{W}^{t}}^T (\boldsymbol{G}^T \mathbf{1}^{L,P})} + {\color{purple} \mu  \oslash (\boldsymbol{H}^t + \alpha)} + \right. \\
    & \left. {\color{orange} 8 \lambda \max (\boldsymbol{H}) + \lambda \boldsymbol{H} \boldsymbol{\Delta}} \Bigr) \right..
    \end{split}
\end{flalign}
To derive this update, we use the fact that ${\color{orange} \nabla r_{\Delta} (\boldsymbol{H}) = \lambda \boldsymbol{\Delta} \boldsymbol{H}}$ and ${\color{orange} \sigma_L = 4}$ since it is the maximimum eigenvalue of the two dimensional Laplacian matrix $\Delta$. Additionally, we note that ${\color{purple} \frac{\partial}{ \partial h_{kp}} \mu\log \left( h_{kp} + \alpha \right) = \frac{\mu}{ h_{kp} + \alpha }}$. Since there is no simplex constraint, to enforce the positivity constraint $\boldsymbol{H}\geq \epsilon$, we use \cite[eq. 22]{perraudin2024efficient} with $\boldsymbol{e}=0$ and obtain the update \eqref{eqn:update_H}.

Similarly, for $\boldsymbol{W}$, we obtain from \cite[eq. 16,17,18 of Proposition 1]{perraudin2024efficient}:

\begin{flalign}
    \begin{split}
        \boldsymbol{W}^{t+1} & = \boldsymbol{W}^t \odot \left( {\color{green} \boldsymbol{G}^{T}\left(\boldsymbol{Y}\oslash \boldsymbol{GW}^{t}\boldsymbol{H}^{t}\right){\boldsymbol{H}^{t}}^{T}}\right) \\ 
        & \oslash \left( {\color{green}\boldsymbol{G}^{T}\mathbf{1}^{L,P}{\boldsymbol{H}^{t}}^{T}}\right)
    \end{split}
\end{flalign}
The simplex constraint from \eqref{eqn:full_loss} reads $\sum_{m \leq M-2} \boldsymbol{W}_m = \boldsymbol{1} = \boldsymbol{u}^T \boldsymbol{W}$ where $\boldsymbol{u}= (u_{m}) \in \mathbb{N}^{M}$ is defined as a vector with $u_m = 1, m \in {1, \dots, M'}$, $u_{M-1} = 0$, and $u_{M} = 0$. We obtain the update \eqref{eqn:update_W} by using \cite[eq. 22]{perraudin2024efficient} with $\boldsymbol{e}=\boldsymbol{u}$. Following the methodology of \cite{perraudin2024efficient}, we obtain the value of the Lagrange multiplier $\boldsymbol{\nu}$ using dichotomy to solve Eq.~\eqref{eqn:nu_eqn}.

Finally, following the iterative application of these updates in \cite[Algorithm 2]{perraudin2024efficient} leads to our proposed Algorithm \ref{alg:espm}.


\bibliography{apssamp}

\end{document}